\documentclass[a4paper,10pt,twocolumn,final,oneside,conference]{IEEEtran}
\usepackage[left=1.60cm,right=1.60cm,top=1.90cm,bottom=2.54cm]{geometry}
\usepackage{cite}
\usepackage{amsmath,amssymb,amsfonts}
\usepackage{algorithmic}
\usepackage{graphicx}
\usepackage{textcomp}
\usepackage{xcolor}
\usepackage{multicol}
\usepackage{balance}
\def\BibTeX{{\rm B\kern-.05em{\sc i\kern-.025em b}\kern-.08em
    T\kern-.1667em\lower.7ex\hbox{E}\kern-.125emX}}
    
\usepackage{tikz}

\IEEEoverridecommandlockouts

\newcommand\copyrighttext{%
  \footnotesize
  \centering Paper submitted to IEEE EEEIC 2022}
\newcommand\copyrightnotice{%
\begin{tikzpicture}[remember picture,overlay]
\node[anchor=south,yshift=0pt] at (current page.south) {\setlength{\fboxrule}{0pt}\fbox{\parbox{\dimexpr\textwidth-\fboxsep-\fboxrule\relax}{\copyrighttext}}};
\end{tikzpicture}%
}

\begin{document}

\title{Characterization of electric consumers through an automated clustering pipeline}

\author{\IEEEauthorblockN{F. Soldan, A. Maldarella,\\
G. Paludetto, E. Bionda}
\IEEEauthorblockA{T\&D Technologies Department\\
RSE S.p.A.\\
francesca.soldan@rse-web.it}
\and
\IEEEauthorblockN{F. Belloni}
\IEEEauthorblockA{\\%
Operation Planning Department\\
Unareti S.p.A.\\
federico.belloni@unareti.it}
\and
\IEEEauthorblockN{S. Grillo}
\IEEEauthorblockA{\\%
DEIB\\
Politecnico di Milano\\
samuele.grillo@polimi.it}}

\IEEEaftertitletext{\copyrightnotice\vspace{0.2\baselineskip}}

\maketitle

\begin{abstract}
Clustering analysis of daily load profiles represents an effective technique to classify and aggregate electric users based on their actual consumption patterns. Among other purposes, it may be exploited as a preliminary stage for load forecasting, which is applied in the same way to consumers in the same cluster. Several clustering algorithms have been proposed and developed in the literature, and the choice of the most appropriate set of clustering parameters is crucial for ensuring reliable results. In this paper, an automated service, suited for repeated clustering analysis, is presented. The pipeline is able to process a generic time series dataset and is easily adjustable to test other clustering input parameters; therefore, it may be utilized to find the best set of parameters with the specific dataset. Moreover, it facilitates repeated characterization on real-time load profiles with the aim of detecting sudden changes of consumers behaviors and variable external conditions, which influence the real power forecasting activity on a short temporal scale.
\end{abstract}

\begin{IEEEkeywords}
automated pipeline, daily load profiles, electric consumption, load forecasting, time series clustering.
\end{IEEEkeywords}

\section{Introduction}
The classification and aggregation of electric users is an important stage in planning and operating the distribution grid. For instance, it may represent a preliminary step for the activity of load forecasting, which is applied in the same way to consumers sharing similar features.

A first approach is to group consumers on the basis of contract information and energy consumption in a given time period \cite{dellagiustina}. However, in this way the grouping represents a snapshot of a given period and does not consider the shape of the energy consumption during time \cite{cerquitelli}. As a matter of fact, demand patterns vary substantially across groups of customers with similar contract characteristics and may be subjected to sudden changes, depending on user behaviors variations.

An alternative solution is to split consumers into groups based on their actual consumption patterns and to frequently repeat the classification step to detect possible changes.

In this context, clustering has been proven to be an effective technique to identify representative load profiles. For example, authors in \cite{yang} used a clustering algorithm to detect energy usage patterns, as a pre-processing step to improve the accuracy of forecasting models. In \cite{bosisio} a performance assessment of clustering methods for load profiles is presented. Authors exploited clustering results to compute some standard load profiles and give a valuable output for planning and operation purposes. In \cite{linden} domestic and small non-domestic electricity customers are classified into several categories based upon different aspects of their consumption patterns. In \cite{Teeraratkul} authors addressed household load curve clustering using a shape-based approach and presented an effective method for estimating when and what devices in a household is used.

Different clustering algorithms are available and the obtained representations of the data are affected by the fixed set of clustering parameters. The choice of the best clustering algorithm, together with the most appropriate set of parameters, is strongly dependent on the specific use case and type of time series to analyze. Therefore, it is necessary to perform repeated clustering analysis to find the most suitable set of parameters for the specific time series dataset.

\begin{figure*}[h]
	\centering
	\includegraphics[width=17cm]{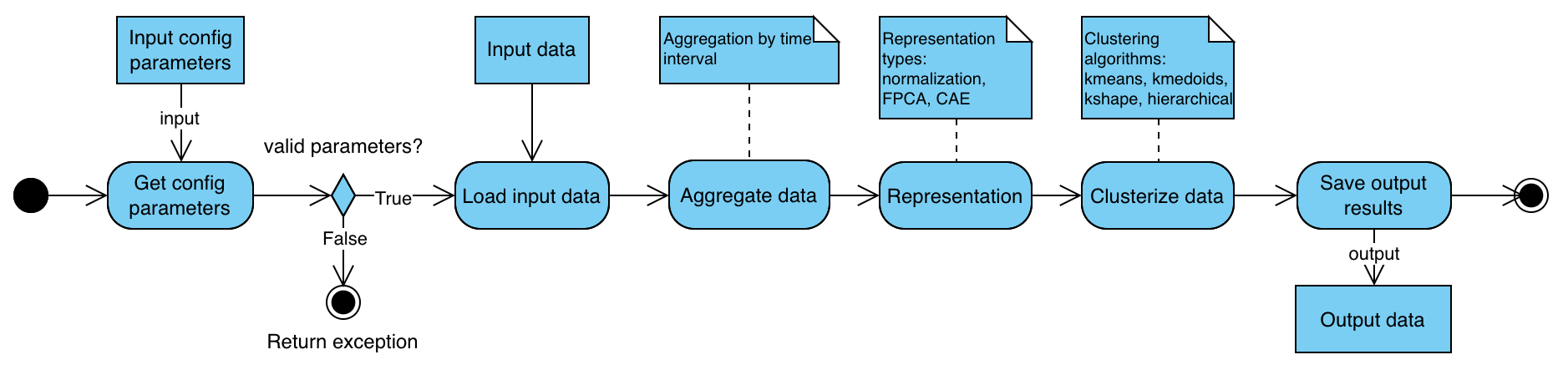}
	\caption{Activity diagram of the implemented process for time series clustering.}
	\label{Fig1}
\end{figure*}

This paper presents an automated clustering pipeline useful for repeated clustering jobs. On the one hand, it enables an easy integration and test of different input configurations; on the other, it facilitates a more frequent aggregation of electric users based on their actual consumption patterns, which is a fundamental prerequisite for more accurate load forecasts.

To test the pipeline, we have considered the real power daily profiles of medium voltage (MV) secondary substations, belonging to the Milan distribution grid, managed by Unareti. Once Unareti will complete the installation of 2G metering systems, the availability of measured customer load profiles will be guaranteed: the analysis of measured profiles (and not of aggregated data at the secondary substation level) will be of major interest for the scope of this work.\IEEEpubidadjcol

\section{Clustering service description}
Figure \ref{Fig1} illustrates the activity diagram of the clustering pipeline, implemented on the cloud platform Databricks. The job is activated through a front-end interface and is composed of the stages detailed below. It is noteworthy that the implemented pipeline is easily adjustable to test other representation modes, clustering algorithms, measure types or validity indexes.

\subsubsection{Import input parameters}
Input parameters are read and validated. They define the desired configuration and include for instance the time aggregation of data, the type of normalization, the possible application of other representation functions and the clustering algorithm.

\subsubsection{Import input dataframe}
Time series dataframe is imported from S3 Amazon Web Services data storage. In this step, it is necessary to tell which column is the x-variable (i.e., the time stamps) and which ones are values of the time series (the y-variable). Along with these pieces of information, it is also vital to tell which column includes the labels. This allows to separate the curves to be clusterized (in our use case this is the column containing secondary substations' names). In this way, it is ensured a standard procedure, applicable to a generic time series dataset.

\subsubsection{Data aggregation}
Data are aggregated over the chosen time interval, using a sum or a mean. Moreover, missing data are estimated using an interpolation method.

\subsubsection{Representation}
Time series data are normalized using z-score, mean or min-max normalizations. In addition, representation-based methods can be applied to reduce the dimension of time series, extract the underlying structure of data and remove noise or other effects. The pipeline includes two representation modes: convolutional auto-encoder (CAE) artificial neural network and functional principal component analysis (FPCA).

CAEs are convolutional operators intended to learn compressed representations of the data through non-linear transformations and therefore perform feature extractions. They are suited to temporal data, as they allow local shift-invariance and capture the shape of time series. This representation step has been already tested in~\cite{CAE}, as a preliminary phase for clustering electric consumption time series.

FPCA is a statistical method to represent functional data in an orthonormal basis of the Hilbert space, that consists of the eigenfunctions of the covariance operator \cite{FPCA}. Selecting a number of $p$ principal components (less than the time series original length) is enough to represent the functional data with a certain level of precision, while providing the benefit of dataset reduction.
\subsubsection{Clustering}
The input dataset is split into groups, using one of the following algorithms: k-means \cite{kmeans}, k-medoids \cite{kmedoids}, agglomerative hierarchical \cite{hier} and k-shape \cite{kshape}.
k-means and k-medoids are the most popular clustering algorithms in the literature and minimize the distance from the cluster center, which is respectively computed as arithmetic mean and as the medoid. Agglomerative hierarchical clustering is another traditional method, which groups input data over various scales creating a cluster tree or dendrogram. Instead, k-shape is a novel method, able to identify shape patterns in time series data, adopting a normalized cross-correlation algorithm.

Together with the choice of the algorithm, it is necessary to select all clustering required parameters, as the distance type (euclidean, k-shape distance or dynamic time warping) and the number of clusters $k$. In particular, the optimal number of clusters should be adjusted in advance, performing the clustering session for a wide range of $k$ values. The choice of the most suitable number of clusters is done using the ``elbow method''~\cite{elbow} or computing the associated cluster validity indexes (CVIs) for each session.

Examples of established CVIs are the Silhouette \cite{sil}, Davies-Bouldin \cite{davies} and Cali\'{n}ski-Harabasz \cite{calinski} scores, which are also computed at the end of the clustering phase to evaluate its goodness. The lower the Davies-Bouldin score is and the higher the silhouette and Cali\'{n}ski-Harabasz indexes are, the better the clustering analysis is.
\subsubsection{Save output files}
Output files are saved in the data storage and include a json file with the original time series labels associated to the belonging clusters and a json file with the cluster validity scores. Moreover, an html file with the graphical representation of time series belonging to each cluster is also saved.

\begin{figure}[h]
	\centering
	\includegraphics[width=8.5cm]{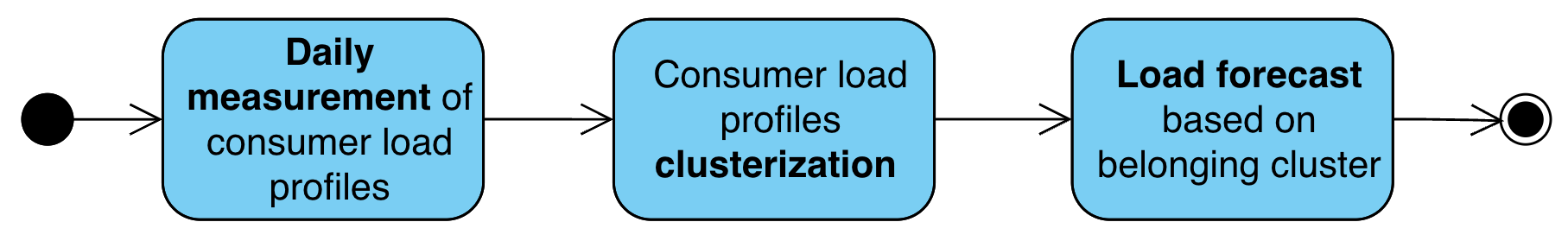}
	\caption{Activity diagram of the load forecasting pipeline which will integrate the clustering service.}
	\label{Fig2}
\end{figure}

The described job will be integrated in a load forecasting pipeline, where clustering is performed to aggregate electric users with similar behaviors, for which the same load forecasting approach is applied (Figure \ref{Fig2}).

\section{Test case for pipeline validation}
\begin{table*}[t]
\caption{Cluster validity indexes (Silhouette (S) score, Davies-Bouldin (DB) score, Cali\'{n}ski-Harabasz (CH) score) and execution times for each combination of parameters (type of algorithm, distance, representation, normalization and processor). The chosen number of clusters is 4. \label{table1}}
\begin{center}
\begin{tabular}{ |c|c|c|c|c|c|c|c|c| }
 \hline
\textbf{Algor.}	&	\textbf{Dist.}	&	\textbf{Rappr.}	&	\textbf{Norm.}	&	\textbf{S score}	&	\textbf{DB score}	&	\textbf{CH score}	&	\textbf{t (s)}	&	\textbf{Processor}	\\
\hline
k-means	&	euclid	&	norm	&	z-score	&	0.31	&	1.04	&	1,190.55	&	26	&	CPU	\\
k-means	&	euclid	&	norm	&	z-score	&	0.31	&	1.03	&	1,190.29	&	20	&	GPU	\\
k-means	&	euclid	&	norm	&	mean	&	0.31	&	1.07	&	1,191.53	&	27	&	CPU	\\
k-means	&	euclid	&	norm	&	min-max	&	0.27	&	1.20	&	1,128.61	&	23	&	CPU	\\
k-shape	&	k-shape-dist	&	norm	&	z-score	&	0.28	&	1.09	&	911.58	&	30	&	CPU	\\
k-shape	&	k-shape-dist	&	norm	&	mean	&	0.31	&	1.65	&	639.07	&	30	&	CPU	\\
k-shape	&	k-shape-dist	&	norm	&	min-max	&	0.18	&	2.41	&	609.65	&	70	&	CPU	\\
k-medoids	&	euclid	&	norm	&	z-score	&	0.19	&	1.40	&	1,033.39	&	24	&	CPU	\\
k-medoids	&	euclid	&	norm	&	mean	&	0.23	&	1.35	&	1,089.40	&	20	&	CPU	\\
k-medoids	&	euclid	&	norm	&	min-max	&	0.25	&	1.22	&	1,099.96	&	23	&	CPU	\\
hierarchical	&	euclid	&	norm	&	z-score	&	0.51	&	1.01	&	75.28	&	23	&	CPU	\\
hierarchical	&	euclid	&	norm	&	mean	&	0.37	&	1.00	&	94.66	&	23	&	CPU	\\
hierarchical	&	euclid	&	norm	&	min-max	&	0.36	&	1.10	&	60.13	&	20	&	CPU	\\
k-means	&	euclid	&	fpca*	&	z-score	&	0.32	&	1.04	&	1,088.31	&	549	&	CPU	\\
k-means	&	euclid	&	fpca*	&	z-score	&	0.31	&	1.04	&	1,088.31	&	375	&	GPU	\\
k-means	&	euclid	&	cae**	&	z-score	&	0.45	&	0.75	&	3,098.55	&	225	&	GPU	\\
\hline
\multicolumn{9}{|c|}{* For FPCA considered 3 principal components.}\\
\multicolumn{9}{|c|}{** CAE trained with 128 epochs and batch size 32. Considered 3 principal components.} \\
\hline
\end{tabular}
\end{center}
\end{table*}

\subsection{Input data}
To test the developed service, we considered the measurements, taken every 15 minutes, of the real power of Milan MV secondary substations. Data cover the period from January to October 2017, for a total number of 2,103 secondary substations, and were made available by Milan's distribution system operator (Unareti S.p.A.). Contractual information about numbers of residential and non-residential contracts per substation were exploited to validate the results.

Raw data were subjected to a pre-elaboration phase, before entering the pipeline, with the aim of computing real power profiles.
We considered three different levels of detail:
\begin{enumerate}
\item secondary substations' daily profiles, computed as the average over each daily time step using all data from January to October 2017;
\item secondary substations' daily profiles, computed as the average over each daily time step separately for different months;
\item secondary substations' daily profiles, measured in specific days of the year (and not obtained through a mean operation).
\end{enumerate}

The three different temporal scales have been chosen to understand how switching from average to measured daily curves affects the results in terms of secondary substations grouping. In particular, results with measured profiles in specific days show how the clustering pipeline could be utilized to study similar profiles and their belonging clustering in nearly real-time.

\subsection{Clustering of MV secondary substations profiles}
\subsubsection{Average real power profiles over the entire year}
\begin{figure}[!b]
	\centering
	\includegraphics[width=4cm]{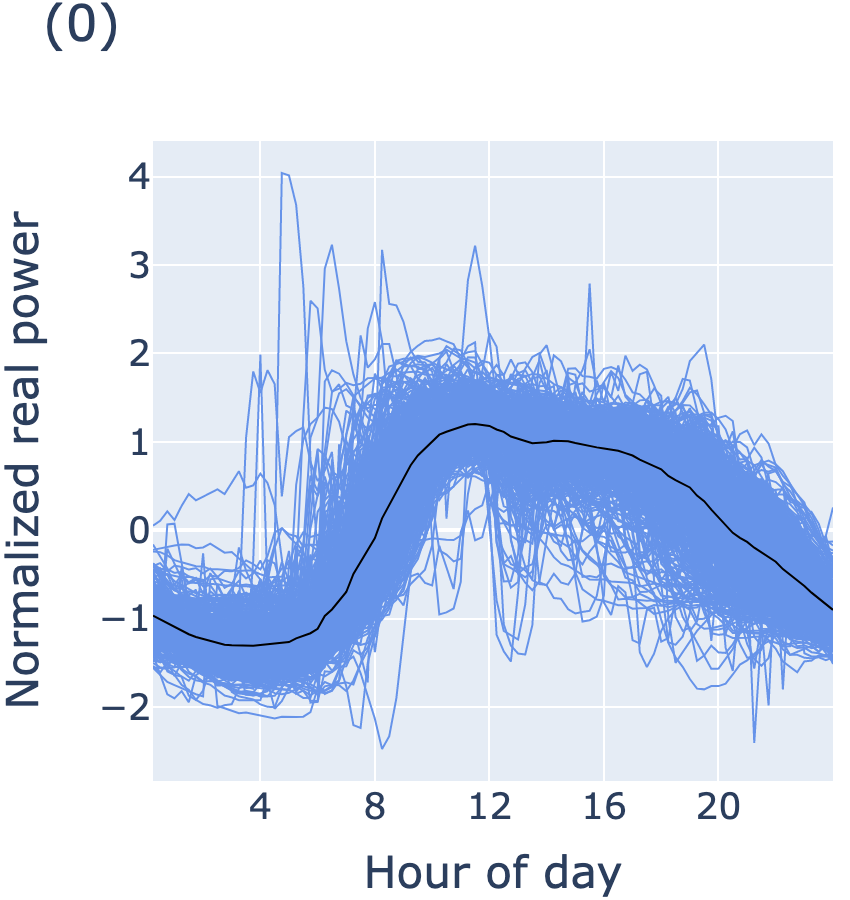}
	\includegraphics[width=4cm]{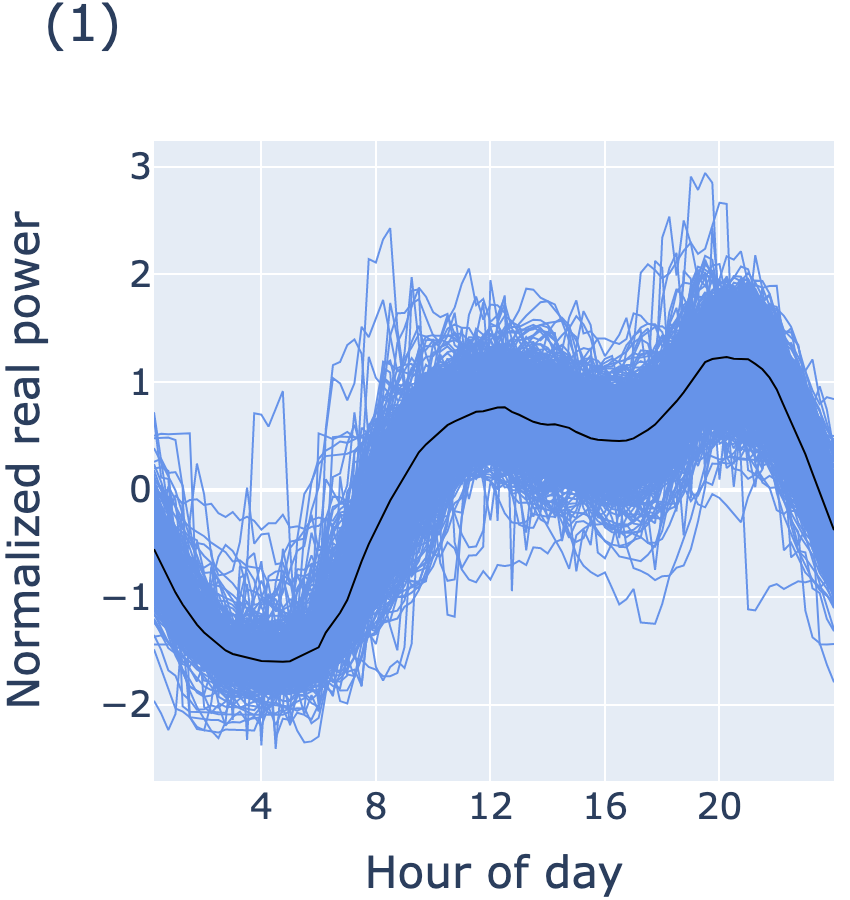}\\
	\includegraphics[width=4cm]{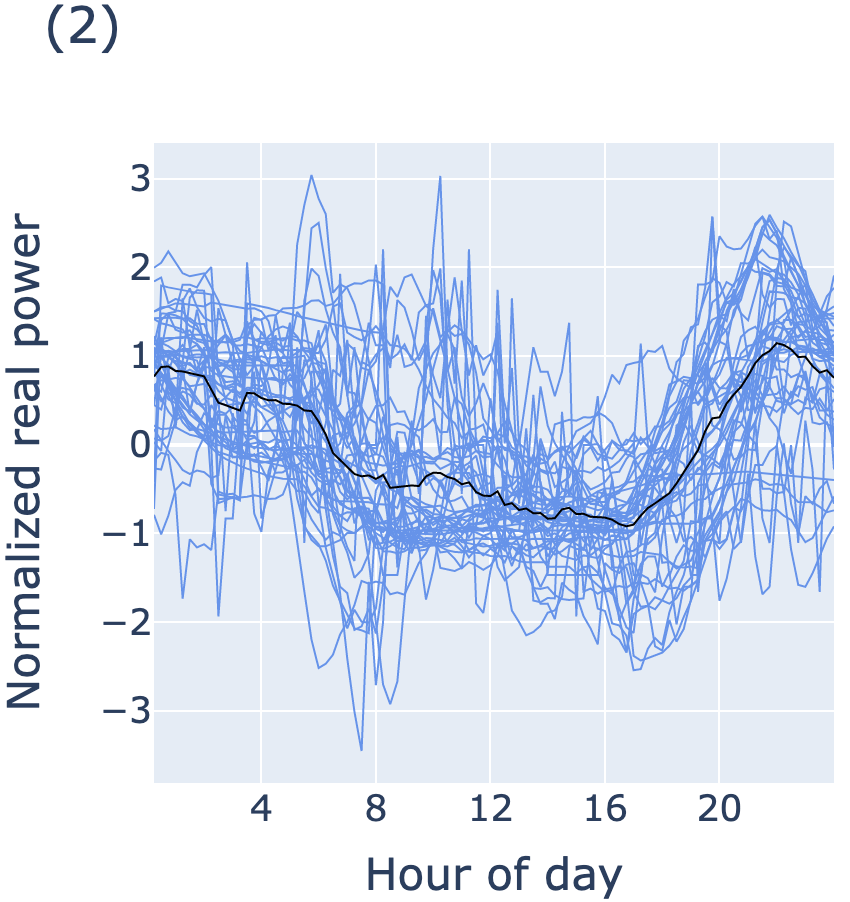}
	\includegraphics[width=4cm]{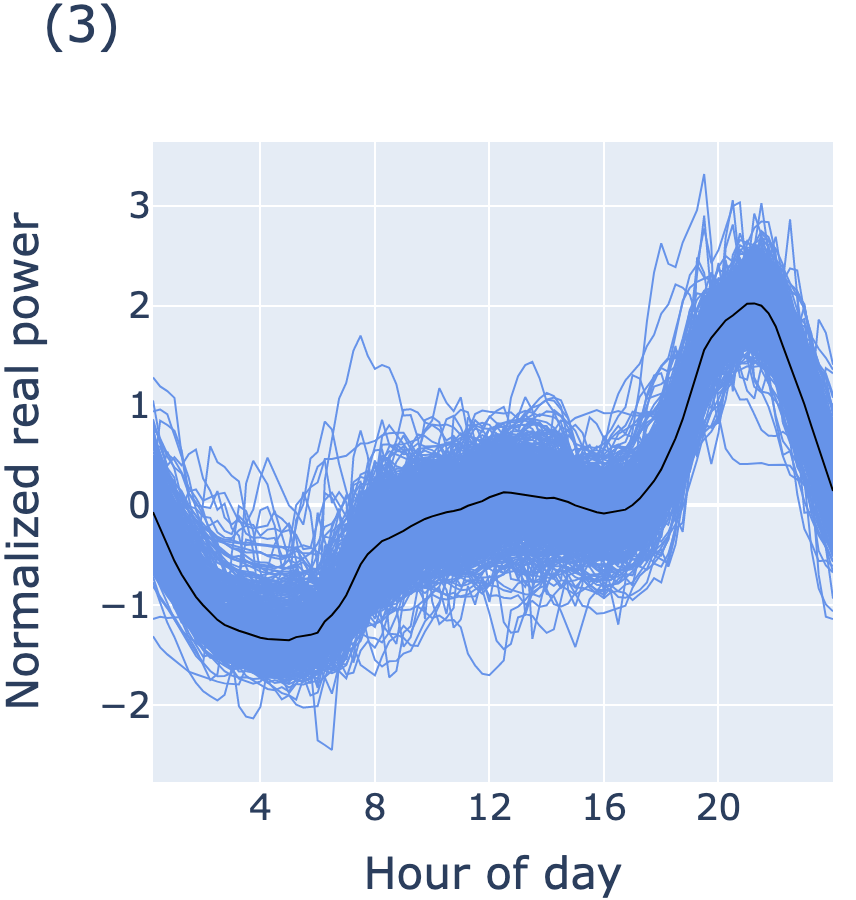}
	\caption{Grouping of average daily profiles of Milan MV secondary substations over the year 2017 (from January to October). The chosen clustering algorithm is k-means, with euclidean distance and z-score normalization. The black curve represents the cluster centroid.}
	\label{Fig3}
\end{figure}
This first use case was exploited to test various combinations of parameters and find the best set for the specific dataset. We considered different combinations of clustering algorithms, distance measures, normalization types and representation modes, and analyzed the results in terms of CVIs (Silhouette, Davies-Bouldin and Cali\'{n}ski-Harabasz scores) and of execution times. Moreover, two different Databricks sets of computation resources and configurations were tested: the first one supplied with a CPU processor, the second one with a GPU processor.

In Table~\ref{table1} we report the results with an optimal clusters number of 4, obtained through a preliminary assessment of the curves of Silhouette scores, Davies-Bouldin scores and through the elbow method. In this step, we do not include results with dynamic time warping because they were not acceptable compared to other metrics (in terms of both execution times and cluster validity indexes).

It is worth noting that the best execution times (in the range 20-70 seconds) are achieved with the choice of normalization as the only representation, even if just CPU is used as processor. In this case, k-means with euclidean distance and z-score normalization is the best performing algorithm. The optimal values of silhouette score and Davies-Bouldin score of hierarchical clustering are indeed determined by the unbalanced number of cluster elements: two clusters include the most load profiles, while the others identify outliers.

The use of CAE neural network as representation improves all cluster validity indexes, at the cost of greater execution time. When the clustering job is not performed in real-time it may be a convenient solution opting for this normalization type, to obtain better results. Instead, the choice of FPCA does not come with a better clusterization, but causes a greater execution time.

Figure \ref{Fig3} shows the curves for the four obtained clusters, with k-means algorithm, euclidean distance and z-score normalization. Through the evaluation of number of contracts per category, we classified clusters (1) and (2) with a residential prevalence, as respectively 72\% and 93\% of elements have a fraction between $0.50$ and $0.75$ of residential contracts. Cluster (3) has a great influence of non-residential contracts (40\% of elements have a fraction between $0.75$ and $1$ of non-residential contracts), while cluster (0) has hybrid features.

Finally, this first data aggregation for the evaluation of daily load profiles brings out general cluster features, linked to the residential or non-residential contracts prevalence. Moreover, this first use case has been exploited to find the optimal parameter configuration with available data.

\subsubsection{Average real power profiles over different months}

\begin{figure}[!h]
	\centering
	\includegraphics[width=8.5cm]{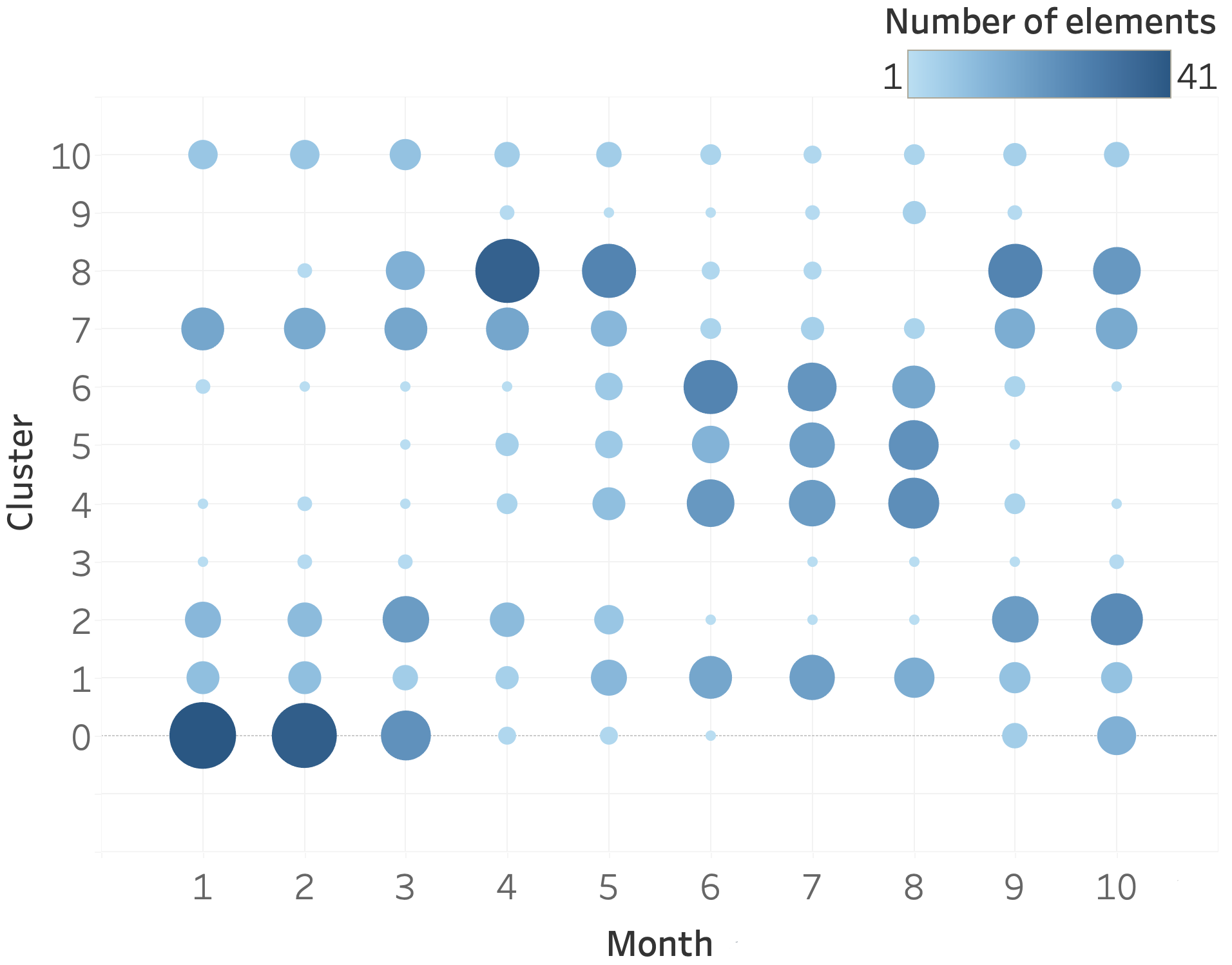}
		\caption{Number of daily profiles in each cluster, for the month from January to October 2017. The circle size is proportional to the number of elements. The chosen clustering algorithm is k-means with euclidean distance and z-score normalization.}
	\label{Fig4}
\end{figure}

\begin{figure}[!h]
	\centering
	\includegraphics[width=4cm]{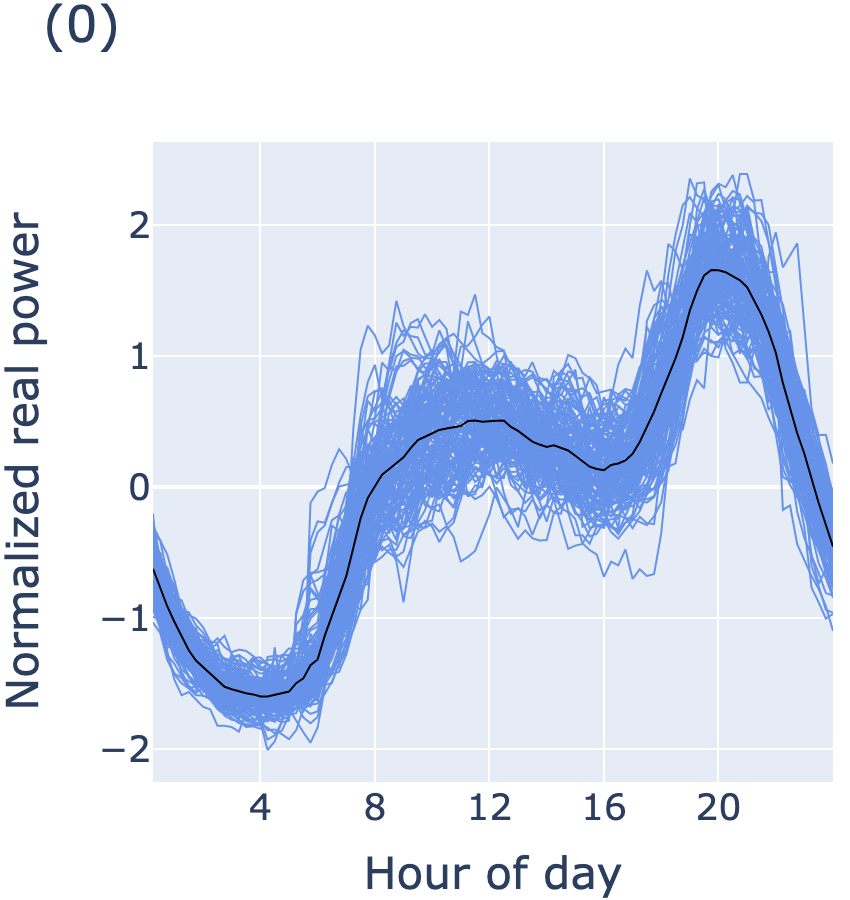}
	\includegraphics[width=4cm]{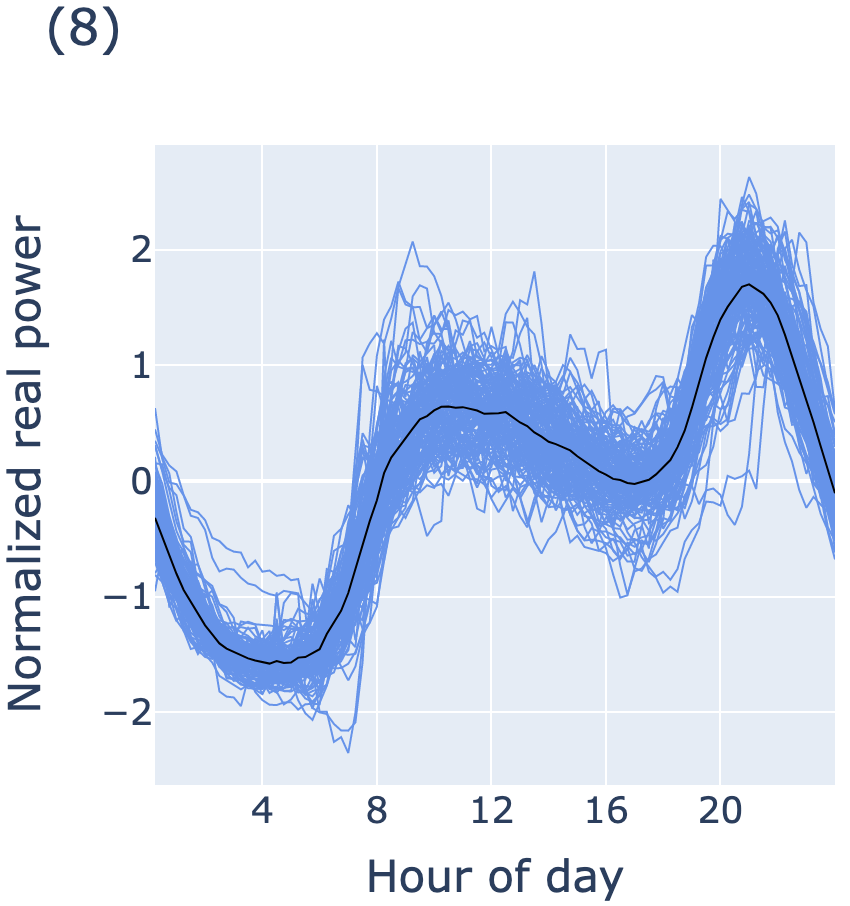}\\
	\vspace{0.2cm}
	\includegraphics[width=4cm]{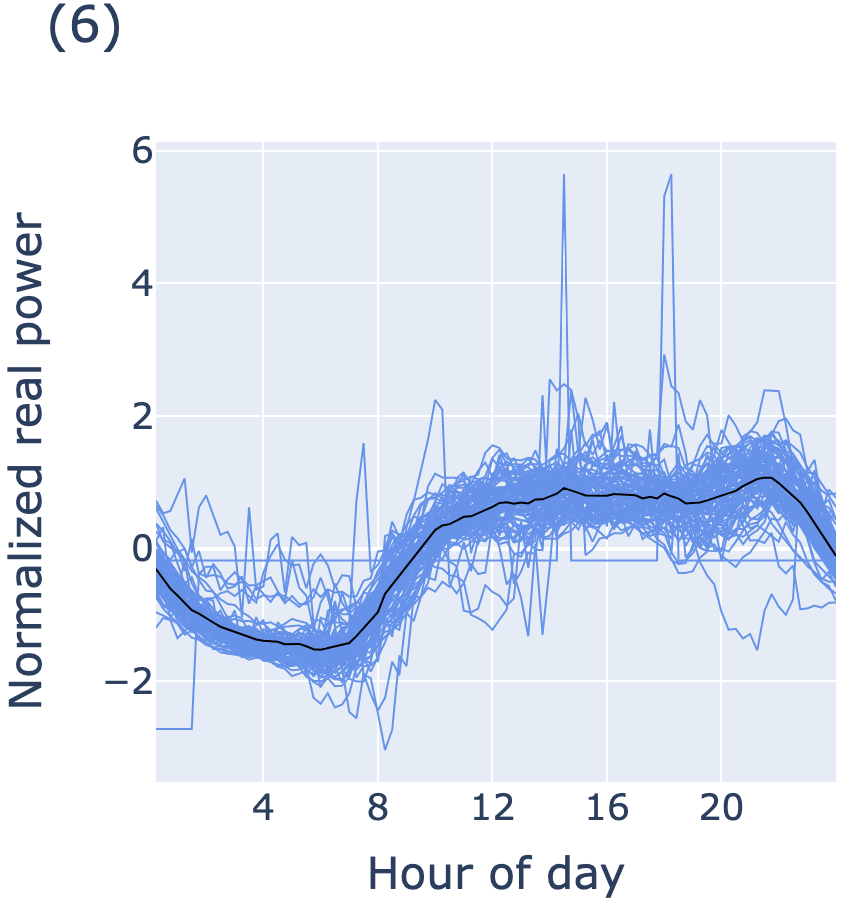}
	\caption{Examples of clusters with a winter (0), spring/autumn (8) and summer (6)  typical traits. The use case is clustering of average daily profiles over different months for Milan MV secondary substations. The chosen clustering algorithm is k-means, with euclidean distance and z-score normalization. The black curve represents the cluster centroid.}
	\label{Fig5}
\end{figure}

For this second case, we selected the 96 secondary substations with the highest data availability in 2017. For each of them, we computed the average daily profile over the months from January to October 2017. The total of 960 curves was ingested by the clustering pipeline. The scope of the analysis was to understand whether a secondary substation stays in the same cluster for the entire year or its profile undergoes shape variations resulting in a shift from one cluster to another.

After a preliminary assessment, we identified an optimal clusters number of 11, with k-means algorithm, euclidean distance and z-score normalization. The results attest a seasonal pattern of some clusters: cluster 0 comprises a greater number of winter time series, cluster 8 is more populated in spring and autumn, while clusters 4, 5 and 6 have a summer typical trait (Figure~\ref{Fig4}). Figure \ref{Fig5} shows the profiles for clusters 0, 8 and 6: in winter (0), spring and autumn (8) it is evident the typical daily profile with a first peak in the morning, followed by a valley and a second peak in the evening; the adoption of air conditioning in summer months (6) causes a curve flattening in the central hours of the day. Moving from a yearly average to a monthly average it is therefore possible to detect seasonal behaviors of MV secondary substations.

\subsubsection{Real power profiles measured in specific days}
\begin{figure}[!b]
	\centering
	\includegraphics[width=4cm]{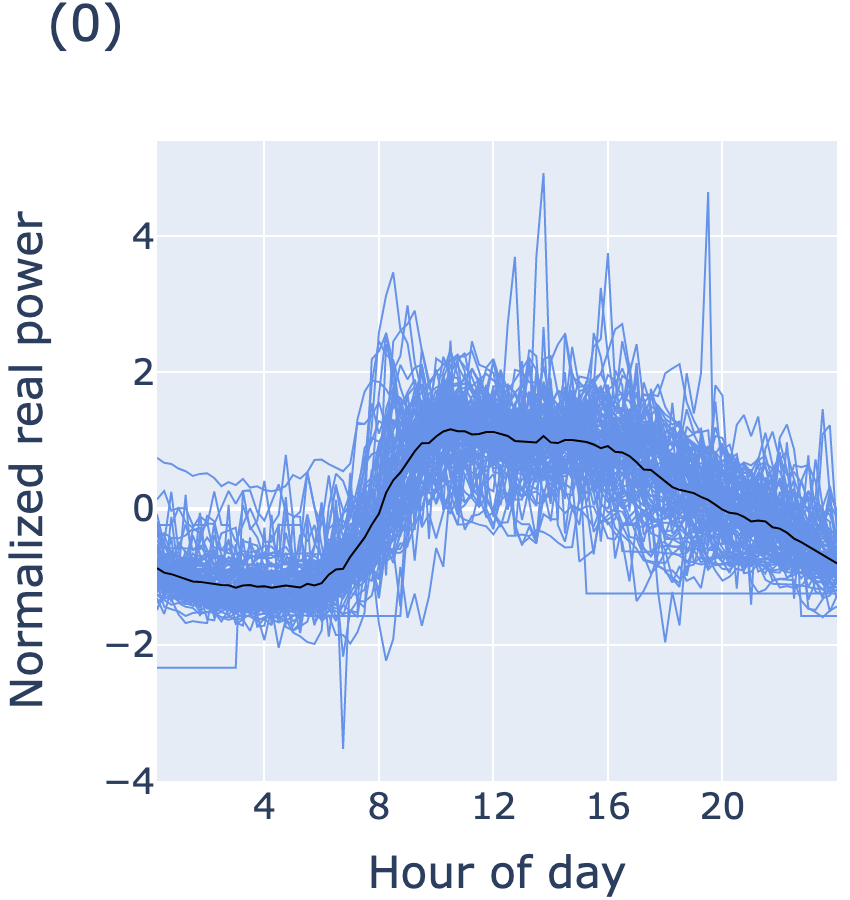}
	\includegraphics[width=4cm]{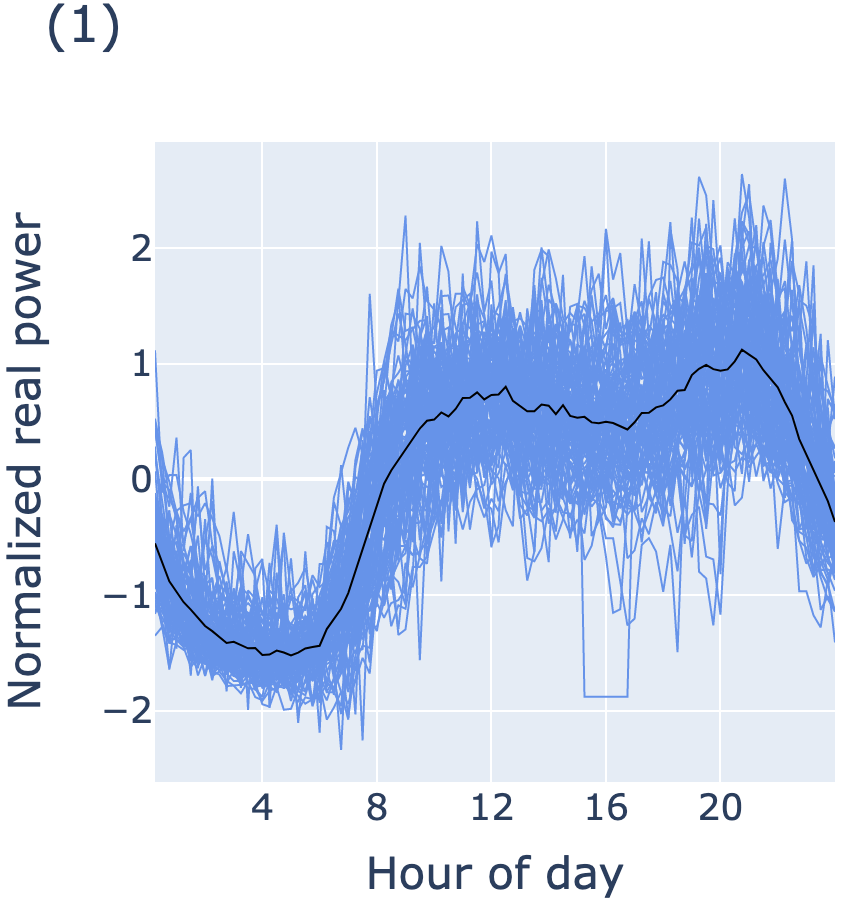}\\
	\includegraphics[width=4cm]{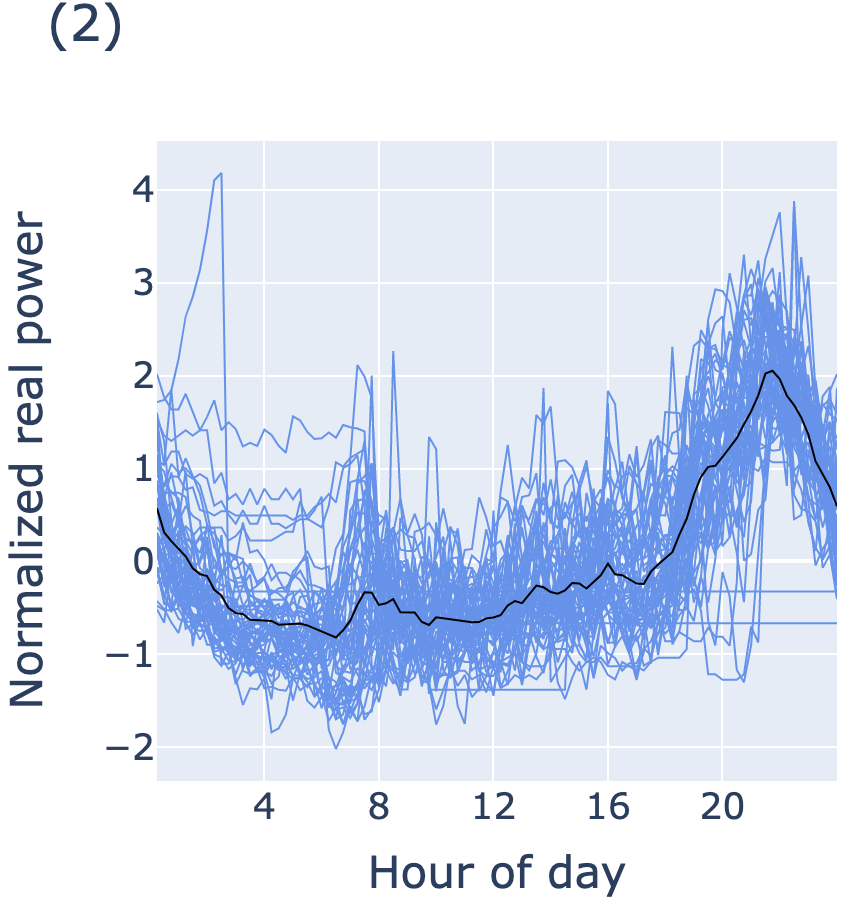}
	\includegraphics[width=4cm]{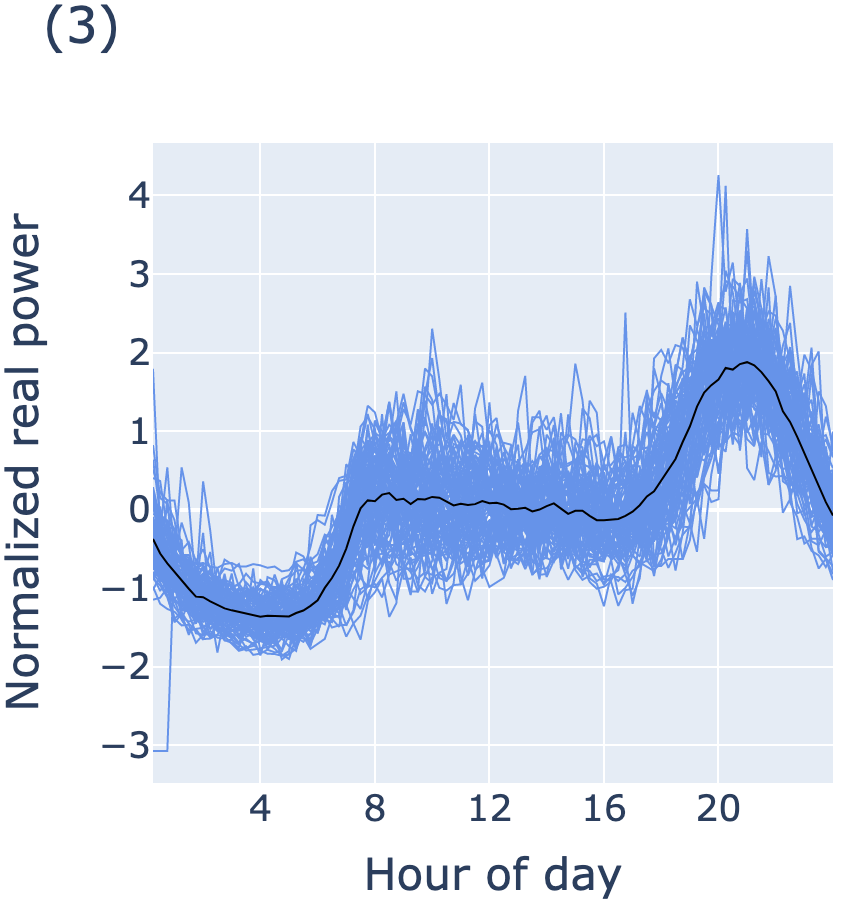}
	\caption{Grouping of daily profiles in specific days (18 Jan. 2017, 19 Apr. 2017, 19 Jul. 2017 and 18 Oct. 2017) for the Milan MV secondary substations. The chosen clustering algorithm is k-means, with euclidean distance and z-score normalization. The black curve represents the cluster centroid.}
	\label{Fig6}
\end{figure}
Finally, to draw closer to a real-time utilization of the developed pipeline, we have selected the measured load profiles in 4 specific days of the year for the 96 secondary substations with the highest data availability (for a total of 384 curves). The selected days are 18 Jan. 2017, 19 Apr. 2017, 19 Jul. 2017, and 18 Oct. 2017, corresponding to four Wednesdays in different seasons of the year. Figure~\ref{Fig6} shows the 4 clusters obtained with k-means, euclidean distance and z-score normalization.

\begin{figure}[!h]
	\centering
	\includegraphics[width=7cm]{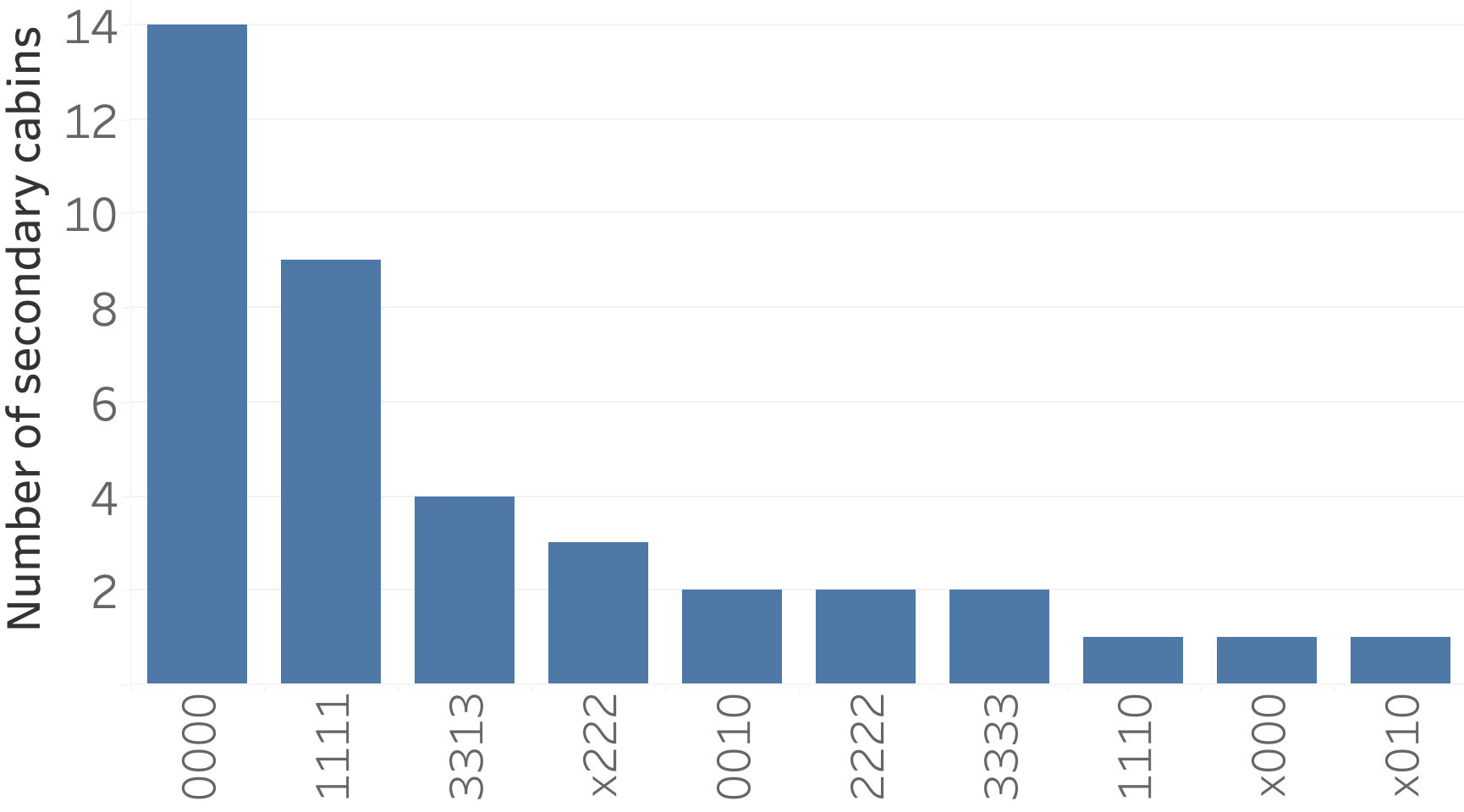}
		\caption{Number of secondary for each label, indicating the belonging clusters in the 4 selected days. "x" indicates absence of data for that day of the year.}
	\label{Fig7}
\end{figure}

Each secondary substation has been labeled with a 4-digit code with pattern WXYZ, where W, X, Y, Z are respectively the numbers indicating the associated cluster on 18 Jan. 2017, 19 Apr. 2017, 19 Jul. 2017, and 18 Oct. 2017. In this way, we found the following common behaviors (Figure~\ref{Fig7}):
\begin{itemize}
\item most secondary substations remain in all seasons in the same cluster (cluster 0 or 1);
\item four secondary substations shift to cluster 1 in summer, while staying in cluster 3 in the remaining seasons. In winter, spring and autumn it is evident the typical curve with two peaks (with the highest peak in the evening hours), whereas in summer the curve undergoes a flattening in the central hours of the day.
\end{itemize}

Finally, this last clustering use case has brought out the possible temporal moving of secondary substations in different clusters, according to the specific day of the year. We have also detected prevailing behaviors, common to specific types of secondary substations.
The clusterization of measured daily curves (not subjected to mean operations) enables the real-time analysis of daily profiles of secondary substations.

\section{Conclusion}
In this paper we presented an automated service performing time series clustering analysis. The implemented pipeline offers a standard procedure, applicable to a generic time series dataset and easily adjustable to include other representation modes, clustering algorithms, measure types or cluster validity indexes. The pipeline enables to easily perform repeated clustering analysis, which is necessary to find the most appropriate set of parameters depending on the specific time series dataset. At this stage, we were able to work on aggregated load profiles at the secondary substation level of Milan MV distribution grid, which were used to validate the pipeline. For this dataset, repeated clustering analyses have shown that the best performing clustering algorithm is k-means, with euclidean distance and z-score normalization. In future activities, we will test the service with data from 2G metering systems, actually being installed on the Milan distribution grid.

Moreover, an automated pipeline facilitates a more frequent characterization of electric users, based on their actual consumption patterns. The clustering activity may be performed every day, to detect possible load profile shape variations, depending on sudden changes of consumers behaviors or variable external conditions.
In the presented use case, real power daily profiles with three different levels of details were considered: yearly real power profiles, monthly real power profiles and real power profiles measured in specific days. In particular, the clusterization of load curves measured in specific days enables a real-time analysis of profiles and therefore the detection of possible behavior variations of consumers on a shorter temporal scale.  In the following activities, the processing chain will be extended, with the integration of a load forecasting job: the clustering service will be a preliminary phase, necessary to find groups of users sharing similar features, for which the same load forecasting approach will be applied.

\section*{Acknowledgement}
This work has been financed by the Research Fund for the Italian Electrical System in compliance with the Decree of Minister of Economic Development April 16, 2018.

\balance



\begin{thebibliography}{10}
\providecommand{\url}[1]{#1}
\csname url@samestyle\endcsname
\providecommand{\newblock}{\relax}
\providecommand{\bibinfo}[2]{#2}
\providecommand{\BIBentrySTDinterwordspacing}{\spaceskip=0pt\relax}
\providecommand{\BIBentryALTinterwordstretchfactor}{4}
\providecommand{\BIBentryALTinterwordspacing}{\spaceskip=\fontdimen2\font plus
\BIBentryALTinterwordstretchfactor\fontdimen3\font minus
  \fontdimen4\font\relax}
\providecommand{\BIBforeignlanguage}[2]{{%
\expandafter\ifx\csname l@#1\endcsname\relax
\typeout{** WARNING: IEEEtran.bst: No hyphenation pattern has been}%
\typeout{** loaded for the language `#1'. Using the pattern for}%
\typeout{** the default language instead.}%
\else
\language=\csname l@#1\endcsname
\fi
#2}}
\providecommand{\BIBdecl}{\relax}
\BIBdecl

\bibitem{cerquitelli}
T.~Cerquitelli, G.~Chicco, E.~Di~Corso, F.~Ventura, G.~Montesano, A.~Del~Pizzo,
  A.~M. González, and E.~M. Sobrino, ``Discovering electricity consumption
  over time for residential consumers through cluster analysis,'' in \emph{2018
  International Conference on Development and Application Systems (DAS)}, 2018,
  pp. 164--169.

\bibitem{yang}
J.~Yang, C.~Ning, C.~Deb, F.~Zhang, D.~Cheong, S.~E. Lee, C.~Sekhar, and K.~W.
  Tham, ``k-shape clustering algorithm for building energy usage patterns
  analysis and forecasting model accuracy improvement,'' \emph{Energy and
  Buildings}, vol. 146, pp. 27--37, 2017.

\bibitem{bosisio}
A.~Bosisio, A.~Berizzi, A.~Morotti, B.~Greco, G.~Iannarelli, C.~Moscatiello,
  C.~Boccaletti, and H.~Noriega, ``Performance assessment of load profiles
  clustering methods based on silhouette analysis,'' in \emph{2021 IEEE
  International Conference on Environment and Electrical Engineering and 2021
  IEEE Industrial and Commercial Power Systems Europe (EEEIC / I CPS Europe)},
  2021, pp. 1--6.

\bibitem{linden}
M.~Lindèn, J.~Helbrink, M.~Nilsson, D.~Pogosjan, J.~Ridenour, and A.~Badano,
  ``Categorization of electricity customers based upon their demand patterns,''
  \emph{CIRED, Open Access Proceedings Journal}, vol. 2017(1), pp. 2628--2631,
  2017.

\bibitem{Teeraratkul}
T.~Teeraratkul, D.~O’Neill, and S.~Lall, ``Shape-based approach to household
  electric load curve clustering and prediction,'' \emph{IEEE Transactions on
  Smart Grid}, vol.~9, no.~5, pp. 5196--5206, 2018.

\bibitem{CAE}
G.~Richard, B.~Grossin, G.~Germaine, G.~Hébrail, and A.~Moliner,
  ``Autoencoder-based time series clustering with energy applications,'' 2020,
  arXiv preprint.

\bibitem{FPCA}
D.~Beretta, S.~Grillo, D.~Pigoli, E.~Bionda, C.~Bossi, and C.~Tornelli,
  ``Functional principal component analysis as a versatile technique to
  understand and predict the electric consumption patterns,'' \emph{Sustainable
  Energy, Grids and Networks}, vol.~21, p. 100308, 2020.

\bibitem{kmeans}
B.~H. Juang and L.~R. Rabiner, ``{The segmental K-means algorithm for
  estimating parameters of hidden Markov models},'' \emph{IEEE Transactions on
  Acoustics, Speech, and Signal Processing}, vol. 38(9), pp. 1639--1641, 1990.

\bibitem{kmedoids}
E.~Schubert and P.~J. Rousseeuw, ``{Faster k-Medoids Clustering: Improving the
  PAM, CLARA, and CLARANS Algorithms},'' \emph{Lecture Notes in Computer
  Science (including subseries Lecture Notes in Artificial Intelligence and
  Lecture Notes in Bioinformatics)}, 2019.

\bibitem{hier}
S.~Patel, S.~Sihmar, and A.~Jatain, ``A study of hierarchical clustering
  algorithms,'' in \emph{2015 2nd International Conference on Computing for
  Sustainable Global Development (INDIACom)}, 2015, pp. 537--541.

\bibitem{kshape}
J.~Paparrizos and L.~Gravano, ``k-shape: Efficient and accurate clustering of
  time series,'' \emph{SIGMOD Rec.}, vol.~45, pp. 69--76, 2016.

\bibitem{elbow}
R.~L. Thorndike, ``Who belongs in the family?'' \emph{Psychometrika}, vol.~18,
  pp. 267--276, 1953.

\bibitem{sil}
P.~J. Rousseeuw, ``Silhouettes: A graphical aid to the interpretation and
  validation of cluster analysis,'' \emph{Journal of Computational and Applied
  Mathematics}, vol.~20, pp. 53--65, 1987.

\bibitem{davies}
D.~L. Davies and D.~W. Bouldin, ``A cluster separation measure,'' \emph{IEEE
  Transactions on Pattern Analysis and Machine Intelligence}, vol. PAMI-1,
  no.~2, pp. 224--227, 1979.

\bibitem{calinski}
T.~Caliński and J.~A. Harabasz, ``A dendrite method for cluster analysis,''
  \emph{Communications in Statistics - Theory and Methods}, vol.~3, pp. 1--27,
  01 1974.

\end{thebibliography}
\end{document}